\documentclass[pre,reprint,superscriptaddress,showpacs]{revtex4-1}
\usepackage[latin1]{inputenc}
\usepackage[sort&compress]{natbib}
\usepackage[dvips]{graphicx}
\usepackage{verbatim}
\usepackage{amsmath}
\usepackage{amsfonts}
\usepackage{amssymb}
\usepackage{latexsym}
\newcommand{\beq}{\begin{equation}}
\newcommand{\eeq}{\end{equation}}
\newcommand{\pr}{\sigma}

\newcommand{\man}{\mathcal{M}}

\newcommand{\kb}{\bar k}

\newcommand{\qt}{\tilde q}
\newcommand{\media}[2]{\left< #1 \right>_{#2}}

\begin{document}

\title{Preferential attachment in growing spatial networks}
%%
%%
%\begin{comment}
\author{Luca Ferretti} 
\affiliation{%%
Centre de Recerca en Agrigen\`omica and Departament de Ci\`encia Animal i dels Aliments, Universitat Aut\`onoma de Barcelona, 08193 Bellaterra, Spain}
\author{Michele Cortelezzi}
\affiliation{Dipartimento di Fisica, Universit\`a di Pisa, Largo Bruno Pontecorvo 3, 56127 Pisa, Italy}

\begin{abstract}
We obtain the degree distribution for a class of growing network models on flat and curved spaces. These models evolve by preferential attachment weighted by a function of the distance between nodes. The degree distribution of these models is similar to the one of the fitness model of Bianconi and Barabasi, with a fitness distribution dependent on the metric and the density of nodes. We show that curvature singularities in these spaces can give rise to asymptotic Bose-Einstein condensation, but transient condensation can be observed also in smooth hyperbolic spaces with strong curvature. We provide numerical results for spaces of constant curvature (sphere, flat and hyperbolic space) and we discuss the conditions for the breakdown of this approach and the critical points of the transition to distance-dominated attachment. Finally we discuss the distribution of link lengths.

\end{abstract}

%\end{comment}
\maketitle
\section{Introduction}

Scale-free networks have attracted a wide interest as models for many systems. Their degree distribution can be explained by the mechanism of preferential attachment in growing networks \cite{barabasi1999emergence}. Preferential attachment has been proposed as a realistic effective mechanism of network growth for many real-world examples, from Internet and the World Wide Web to Wikipedia and citation networks. However, preferential attachment alone is not enough to explain the structure of scale-free networks, since many of these networks build also on other variables like %%
fitness \cite{bianconi2001competition,bianconi2001bose} and %%
spatial structure \cite{manna2002modulated,xulvi2002evolving,barthelemy2003crossover,santiago2007emergence,santiago2008extended,flaxman2006,flaxman2007,jordan2010}. 

The fitness model of Bianconi and Barabasi \cite{bianconi2001competition} is the basic example of a network model where the attachment depends not only on the number of links but also on another variable, namely the quality of the nodes. This model predicts multiple power-law scaling in the degree distribution and a phase with Bose-Einstein condensation on nodes with high fitness \cite{bianconi2001bose} and it has been applied to the WWW \cite{pastor2001dynamical,vazquez2002large}, where Google could be an interesting example of an emerging condensate.

Space is another feature that plays an important role in many real-world networks. Examples include the world airport network, which is scale-free \cite{barrat2004architecture} and where distances are relevant \cite{bianconi2009assessing}, but physical and abstract spaces appear in many other transportation networks as well as in communication and social networks. Another biological example is given by brain connectivity in the hippocampus, which depends on the distance in physical space and shows a scale-free degree distribution \cite{bonifazi2009gabaergic} and a short axon length compared to the size of the region. 

It is possible to obtain scale-free networks from models based on on hidden variables even without any preferential attachment \cite{caldarelli2002scale,boguna2003class}. Interestingly, there are several models with power-law degree distributions induced by the space itself (see \cite{barthelemy2010spatial} for an extensive review). In some models, it is the interplay/competition between spatial and network rules that generates to a scale-free distribution \cite{fabrikant2002,berger2003,berger2004,berger2005,d2007emergence}. %\cite{fabrikant2002,berger2003,berger2004,berger2005,d2007emergence}. 
Other models rely only on the spatial structure; as an example, recent models based on static random networks on hidden hyperbolic spaces \cite{serrano2008self,krioukov2008efficient,krioukov2009curvature} have been shown to give rise to scale-free degree distributions and could explain the assortativity and clustering observed for Internet. 
 
An alternative, simpler way to obtain spatial scale-free networks is to embed the nodes in a metric space and modify the preferential attachment rule to includes spatial information. The classic preferential attachment mechanism is based on a connection probability $\Pi_{j\rightarrow i}$ that is linear in the degree of the node $\Pi_{j\rightarrow i}\propto k_i$. If each node $i$ is associated to a position $x_i$ in a metric space, an immediate modification of preferential attachment is given by a multiplicative weight depending on the distance $d_{ij}$ between nodes, i.e. 
\beq
\Pi_{j\rightarrow i}\propto \sigma(d_{ij})k_i\label{eq1}
\eeq
The positions $x_i$ can be randomly chosen from a distribution $\rho(x)$. If all distance are equal (for example, if the space reduces to a single point) then the spatial structure disappears and the model reduces to the Barabasi-Albert model, otherwise there could be a non-trivial interplay between the spatial and network structure. Growing network models in this class have been proposed several times in the literature and have been studied numerically in Euclidean spaces \cite{manna2002modulated,xulvi2002evolving,yook2002modeling,barthelemy2003crossover}. These studies are reviewed in \cite{barthelemy2010spatial}, section IV.D.2. Furthermore, analytical results have been obtained for some models on the sphere \cite{flaxman2006,flaxman2007,jordan2010} and in generic spaces under strong approximations  \cite{santiago2007emergence,santiago2008extended}. However, despite the simplicity of these models, there are no analytical or numerical results for models on generic metric spaces and with generic weights $\sigma(d)$. Our aim is to provide a general analysis of this class of models.

In this paper we focus on networks on flat and curved spaces, the growth of which follows a preferential attachment rule weighted by a function of the distance between nodes as in equation (\ref{eq1}).  %%
Our main result is the derivation of a general expression for the asymptotic degree distribution. The form of the degree distribution is essentially equivalent to the fitness model of Bianconi and Barabasi \cite{bianconi2001competition}, with a fitness distribution that depends on the space and on the connection %%
function $\sigma(d)$ through a selfconsistency equation.
We present some examples where the distribution can be solved explicitly. These examples include homogeneous space and spaces with curvature singularities; in the latter, Bose-Einstein condensation can occur on nodes near a singularity. We also study numerically the distribution for disks in the flat plane, on the sphere and in the hyperbolic plane, showing that in the hyperbolic model there is a long but transient regime of Bose-Einstein condensation. Finally we give a general expression for the distribution of link lengths.

\section{Degree distribution of growing networks on hidden spaces}
We present a general class of models for networks with preferential attachment on a metric space. In these models, each node has a position $x$ randomly chosen from a space $\man$ with probability $\rho(x)$. We call $x_i$ and $k_i$ the position and degree of the $i$th node. At each time step, a node with $m$ links is added to the network and the $m$ links are randomly attached to nodes in the network with a modified preferential attachment rule.

We are mostly interested in the case where each node is assigned a positions on a Riemannian manifold of dimension $D$ (possibly with boundaries and/or singularities) with finite volume, that is, $S=(\man,ds^2)$ where $\man$ is the manifold and $ds^2$ is its metric, that is, the squared distance between the points $x$ and $x+dx$, which is a bilinear form $ds^2=\sum_{i,j}g_{ij}(x)dx_idx_j$ in the infinitesimal displacement $dx$ on the manifold. 
%$ds^2=\sum_{i,j}g_{ij}(x)dx_idx_j$ is the metric of the manifold, that is, the infinitesimal distance between the points $x,x+dx$ induced by a metric tensor $g_{ij}$. 
Specific models in this class were studied in \protect\cite{manna2002modulated,xulvi2002evolving,yook2002modeling,barthelemy2003crossover,flaxman2006,flaxman2007,jordan2010} and a more general model was defined in \protect\cite{santiago2007emergence,santiago2008extended}. However the analysis of \protect\cite{santiago2008extended} is valid only for almost-homogeneous manifolds, as discussed in section \protect\ref{homog}. 

In this work we employ the natural requirement that all the information about hidden variables should come from the metric structure of the manifold. More specifically, we require that nodes are assigned random position on the manifold such that the average number of nodes per unit volume is constant. Therefore the resulting density is simply
$\rho(x)=1/\mathrm{Vol}(\man)$ if expressed in terms of (locally) Euclidean coordinates, or more generally $\rho(x)=\mathcal{J}^{-1}(x)/\mathrm{Vol}(\man)$ where $\mathcal{J}(x)$ is the Jacobian of the change of coordinates to (locally) Euclidean coordinates. However, all the results of this section extend naturally to the case of a generic node density $\rho(x)$ arbitrarily chosen.

Moreover, we require that the preferential attachment rule be modified such that the connection probability from a new node with coordinate $y$ to a node with coordinate $x$ depends on their distance $d(x,y)$, that is, the attachment rule is given by a real positive function $\pr(d)$ such that the probability to attach a link to the $i$th node is
\beq
\Pi(i)=\frac{\pr(d(x,x_i))k_i}{\sum_{j}%%
\pr(d(x,x_j))k_j}
\eeq
where $x$ corresponds to the position of the new node. These models correspond to the heterogenous network models of \cite{santiago2007emergence,santiago2008extended}, but while these papers focus on the almost-homogeneous case, we allow for all spaces and all levels of heterogeneity.

An important quantity related to $\pr(d)$ is its typical length scale $\lambda_\pr\sim \langle d \pr(d)\rangle/\langle\pr(d)\rangle$, which %%
depends explicitly on the manifold $\man$ (and on its boundary, if it exists). Denoting by $R$ the typical length scale of the manifold, the connection function is ``long-range'' if $\lambda_\pr\sim R$ and ``short-range'' if $\lambda_\pr\ll R$. The networks associated with the two cases can have different properties.

A network model on a hidden space is therefore defined by a manifold $\man$ and a connection probability $\pr(d)$. We derive the degree distribution through the rate equation approach introduced in \cite{krapivsky2000connectivity, krapivsky2002}. The equation for the node density $n_k(x)$, which is the average density of nodes with hidden position $x$ and degree $k$, is
\begin{align}
&
n_k(x,t+1)=n_k(x,t)-m\ \times \label{neq} \\ &\times 
\int_\man d^Dy  \frac{\pr(d(x,y))(k n_k(x,t)-(k-1)n_{k-1}(x,t))}{\int_\man d^Dz\ \pr(d(z,y))\sum_{k'=m}^\infty{k'} n_{k'}(z,t)} \rho(y)\nonumber
\label{neq} \end{align}
plus a term $+\delta_{k,m}\rho(x)$ that accounts for the birth of new nodes. 
We assume a linear scaling with time for the quantity
\beq
{\int_\man d^Dz\ \pr(d(z,y))\sum_{k'=m}^\infty{k'}n_{k'}(z,t) }=mC(y)t+o(t)\label{linearscaling}
\eeq
then can rewrite it as 
\beq
n_k(x,t+1)=n_k(x,t)-(k n_k(x,t)-(k-1)n_{k-1}(x,t))\frac{q(x)}{t}
\eeq
where $q(x)$ plays the same role as the (average) fitness of the node \cite{bianconi2001competition,bianconi2001bose} and is defined as 
\beq
q(x)=\int_\man d^Dy  \frac{\pr(d(x,y)) }{C(y)}\rho(y)\label{qdef}
\eeq
The ``fitness'' $q(x)$ is determined by solving asymptotically the above equation
\begin{align}
{n_k(x,t)}&=t\cdot\frac{\rho(x)}{q(x)}\frac{\Gamma(m+q(x)^{-1})\Gamma(k)}{\Gamma(k+1+q(x)^{-1})\Gamma(m)}\nonumber \\
&\simeq t\cdot\frac{\rho(x)}{q(x)m}\left(\frac{k}{m}\right)^{-(1+q(x)^{-1})}\label{nkeq}
\end{align}
and substituting in the definition of $C(y)$ and in (\ref{qdef}) to obtain 
a single functional equation for $q(x)$:
\beq
q(x)=\int_\man d^Dy\  \rho(y)\frac{\pr(d(x,y))}{\int_\man d^Dz\ \rho(z)\frac{\pr(d(z,y))}{1-q(z)}}\label{qeqs}
\eeq
This consistency equation determines $q(x)$ as a function of the global structure of the manifold $S=(\man,ds^2)$, the weight function $\pr(d)$ and the node density $\rho(x)$. 

From the point of view of the degree distribution, this class of models is equivalent to the Bianconi-Barabasi fitness model  \cite{bianconi2001competition,bianconi2001bose}, but in this case the fitness distribution is dynamically determined by $\rho(x)$ and $\pr(d)$ through equation (\ref{qeqs}). The resulting degree distribution is
\beq
p(k)=\int_\man d^Dx\ \frac{\rho(x)}{q(x)m}\left(\frac{k}{m}\right)^{-(1+q(x)^{-1})}\label{pkspace}
\eeq 
so the distribution is a sum of power laws similarly to the fitness model, and for a regular distribution of $x$ and $q(x)$ it typically reduces to a power law with logarithmic corrections \cite{bianconi2001competition}.

Note that if we allow for an arbitrary node density $\rho(x)$, the only relevant information about the manifold is the distance function $d(x,y)$, therefore these results can be applied to any metric space. However from now on we focus on Riemannian manifolds with uniform node density.

We are particularly interested in warped manifolds. A warped manifold is a space $\mathcal{M}$ that is locally a product $\mathbb{R}\times \mathcal{H}$ where $\mathcal{H}$ is an homogeneous space with metric $ds^2_\mathcal{H}$, and whose metric has the form $ds_\mathcal{M}^2=dr^2+f(r)ds^2_\mathcal{H}$. We are interested in spaces with $O(D)$ symmetry, which means that $\mathcal{H}$ is the sphere $S^{D-1}$ with the natural metric $d\Omega^2$. In this case, if there is a radius (say, $r=0$) such that $f(r)\rightarrow 0$ for $r\rightarrow 0$, then $f(r)\sim r^{2}$ for $r\sim 0$ to avoid singularities in the metric. Flat spaces, hyperbolic spaces and hyperspheres can be seen as particular examples of these spaces with $f(r)=r^2$, $f(r)=\sinh^2(\zeta r)/\zeta^2$ and $f(r)=\sin^2(\zeta r)/\zeta^2$ respectively. %%
In these  spaces the $O(D)$ symmetry implies that $q$ and $C$ depend only on $r$ and that $q(r)$ is determined through the equation
\beq
q(r)=\int_{r_{min}}^{r_{max}} ds\  \frac{h(r,s) }{\int_{r_{min}}^{r_{max}} du\ \frac{h(s,u)}{1-q(u)}}\label{qeqw}
\eeq
where $h(r,s)$ is 
\beq
h(r,s)={f(s)^{\frac{D-1}{2}}}\int_{S^{D-1}} d\Omega\ \pr(d(r,\Omega_0;s,\Omega))
\eeq

In the following sections we discuss three analytically solvable cases (homogeneous spaces, balls in flat or hyperbolic spaces with long-range connections, warped throaths) and then the interesting cases of disks on the flat plane, the sphere and the hyperbolic plane, which are the simplest spaces corresponding to constant zero, positive and negative curvature respectively.

\section{Solvable models}

\subsection{Homogeneous spaces}\label{homog}
We consider first the cases where $\man$ is an homogeneous space. This means that there is a symmetry group on $\man$ that leaves the metric (and the measure) invariant and moreover that the orbit of each point in $\man$ under the action of the group is $\man$ itself. The first consequence is that the  volumes $d^Dx\rho(x)$ and the connection probability $\pr(d(x,y))$ are invariant under the action of the symmetry group, as are all quantities build from integrals of functions of the position over the whole manifold $\man$. Then, if the assumption (\ref{linearscaling}) is correct and if all the integrals involved in the definition of $C(x)$ and $q(x)$ are convergent, neither $C(x)$ nor $q(x)$ depend on the position $x$, and therefore integrating both sides of the equation (\ref{qeqs}) by $\int d^Dx\rho(x)$ we obtain $q=1/2$. This means that the degree distribution of the model on homogeneous spaces remains essentially the same as the one of the Barabasi-Albert model
\beq
p(k)\propto k^{-3}\label{pk3}
\eeq
For homogeneous spaces, this result does not depend at all on the form of $\pr(d)$, as long as its integral over the volume is convergent.

The result (\ref{pk3}) was obtained numerically by  Barthelemy \cite{barthelemy2003crossover} for the case of the flat disk in $D=2$ with $\pr(d)=e^{-d/r_c}$ and $r_c$ not too small. In the same paper it was shown that for small $r_c$ the transient regime dominates the dynamics for long times, inducing a cutoff $k_{max}\sim n^{0.1}$ where $n$ is the number of nodes inside a volume of order $r_c^2$. In this section we have shown that the result (\ref{pk3}) is actually general for homogeneous spaces. 

Recently, Jordan \cite{jordan2010} proved a slightly more general result: the degree distribution (\ref{pk3}) is the same for all spaces where the volume of a ball depends only on its radius and not on its position, provided that $\sigma(d)$ has a positive minimum and that the integrals of $\sigma(d)$ and $\sigma(d)^2$ over the whole manifold are finite. This result applies to all homogeneous spaces, confirming the above argument. Note that Jordan's result can be directly obtained from equation (\ref{qeqs}) by substituting $q(x)=1/2$ and interpreting all integrals as Lebesgue integrals, with the only requirement $\int_\man d^Dx\ \sigma(d(x,y))<\infty$. %The possible role of the extra requirement $\int_\man d^Dx\ \sigma(d(x,y))^2<\infty$, which is necessary for $m>1$ in the proof in \cite{jordan2010}, will be discussed later. COMPLETE

In practice, equation (\ref{pk3}) is also true for open balls in homogeneous spaces if $\lambda_\pr\ll R$, where $R$ is the radius of the ball. Consider as an example the hyperbolic disk: if the typical radius $\lambda_\pr$ of the function $\pr(d)$ is much smaller than the radius of the disk, then the disk appears to each node as the (infinite homogeneous) hyperbolic plane. Deviations from $q(r)=1/2$ decay exponentially with the distance from the boundary of the ball. More generally, a rough approximation for almost-homogeneous spaces is to consider $C$ independent on the position on the manifold and therefore obtain the approximate equation \cite{santiago2008extended}
\beq
q(x)=\frac{1}{2}\frac{\int_\man d^Dy\  \rho(y)\pr(d(x,y))}{\int_\man d^Dw \int_\man d^Dz\ \rho(w)\rho(z)\pr(d(w,z))}
\eeq

\subsection{Balls with long-range connections} \label{balld}

We consider balls in flat spaces of high dimension $D\gg 1$. In this limit, most of the nodes lie within a distance $R/D$ from the boundary of the ball. This means that for $D\rightarrow \infty$ almost all the nodes lie approximately on the boundary, that is, on the $(D-1)$-dimensional sphere $S^{D-1}$, which is an homogeneous space with the induced metric. Moreover, if the connection probability grows with the distance, most of the nodes will connect to nodes on the boundary. Therefore, these nodes have $q=1/2$ and it is possible to calculate $q(r)$ for the few nodes in the inside from equation (\ref{qeqw}) by considering an effective distribution $\rho(r)\propto\delta(r-R)$ of nodes. If we choose for example $\pr(d)=d^2$ the result is
\beq
q(r)=\frac{\int_0^\pi d\vartheta\ \sin(\vartheta)^{D-2}(r^2+R^2-2rR\cos(\vartheta))}{\int_0^\pi d\vartheta\ \sin(\vartheta)^{D-2}4R^2(1-\cos(\vartheta))}=\frac{R^2+r^2}{4R^2}
\eeq 
Similar (but more complicated) results can be obtained in the same limit $D\gg 1$ with other functions $\pr(d)$ or with balls in other spaces (hyperspheres, hyperbolic spaces). 

\subsection{Warped throats, condensation and the fitness model}
A natural question about these models is: what happens when equation (\ref{qeqs}) does not admit solutions? The absence of solutions is a sign of the breakdown of the mean-field approximation in  equations (\ref{neq}),(\ref{linearscaling}). This breakdown could appear as condensation of the links on a single node as in the fitness model of Bianconi and Barabasi \cite{bianconi2001bose}. In order to show that this possibility is realized at least in some cases, we naturally embed the condensate phase of the fitness model in a network on a warped geometry. 

\begin{figure}[htb]
\includegraphics[scale=0.5]%%
{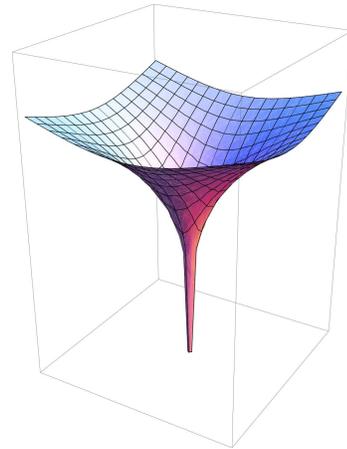}
\caption{\label{throat}Example of a two-dimensional warped throat embedded in tridimensional space.}
\end{figure}

The basic mechanism for condensation can be understood by considering the throat-like geometry illustrated in figure \ref{throat}. Most of the nodes would lie on the upper border of the throat, while only a few nodes would lie near the tip. If the connection function $\sigma(d)$ increases strongly with the distance, many nodes would connect to the node that is closest to the tip, because it would have maximum distance from the border and almost no competitors. In turn, the degree of this node would grow fast and attract links because of preferential attachment. Then, links would condensate on this node.

For a concrete realization of the mapping between our model and the fitness model, we consider networks on warped throats. These geometries are well-known in the context of string theory and particularly of gauge/gravity correspondence \cite{klebanov2000supergravity}. They can be realized as warped spaces $\mathbb{R}^+\times S^{D-1}$ with a metric $ds^2=dr^2+f(r)d\Omega^2$ and a function $f(r)$ that remains very small for a large interval of $r$,  as illustrated in Figure \ref{throat}. In particular, we consider $r\in [0,1]$, $D\gg 1$ and we choose $f(r)=\varepsilon r^k$ and $\pr(d)=d$. Then $\rho(r)=r^{k(D-1)/2}(kD-k+2)/2 $. In the limit $\varepsilon\ll 1$, $kD\gg 1$ almost all the nodes lie at $r\simeq 1$. Moreover, the size of the transverse spherical sections is given by $\sqrt{\varepsilon}$, which is practically negligible, therefore the connection probability can be approximated as $\pr(d_{1,2})=d_{1,2}\simeq|r_1-r_2|$. From the point of view of the degree distribution, it can be further approximated as $\pr(d_{1,2})\simeq 1-r_1$ since $r_2$ would almost surely lie around $1$. If we define $\eta=1-r$, we see that $\pr(d_{1,2})\simeq \eta_1$ and therefore the model corresponds precisely to the Bianconi-Barabasi model with fitness distribution $\rho(\eta)\propto (1-\eta)^{k(D-1)/2}$, that is, in the deep condensate phase \cite{bianconi2001bose}. Then the mean-field approximation breaks down and there is a condensation of links (that is, a finite fraction of the links) on a node near the tip of the throat.

The correspondence between the two models captures well the competition between nodes with higher ``fitness'', that is, the nodes between $r=0$ and $r'\ll 1$. This means that the condensate is not a transient effect but a long-term feature of this geometry. This is related to the presence of a metric singularity on the tip of the throat: in fact a non-singular tip would look locally like $\mathbb{R}^D$, therefore it would contain a finite density of nodes increasing linearly in time and competing asymptotically with the condensate. The competition of nodes with the same fitness would make the degree of the most connected node grow sublinearly ($q<1$) and then the condensate would disappear.

In the above model with $k=2$ %%
it is possible to see explicitly the condensation phase transition  %%
by varying the parameter $\varepsilon$
. This interpolation is possible because in this case $\varepsilon$ has also an easy geometric interpretation: in fact the geometry reduces to a $D$-dimensional hypercone with a conical singularity at the tip $r=0$ with cone angle $\alpha$ and the parameter $\varepsilon$ is a function $\varepsilon=\alpha^2/4\pi^2$ of this angle. For $\varepsilon=1$ we have $\alpha=2\pi$, therefore the space is a flat ball without singularities and the model reduces to the one of the previous section \ref{balld} with a degree distribution $p(k)\sim k^{-3}$. 
On the other extreme, for $\varepsilon\rightarrow 0$ the cone is extremely shallow ($\alpha\rightarrow 0$) and we recover the model above, which is in the deep condensate phase. 
From geometrical reasoning, the critical point for this phase transition in the limit $D\rightarrow\infty$ lies around %%
$\varepsilon_c\simeq 16\arcsin^2(1/4)/\pi^2\sim 0.1$.

It is not clear if there can be condensation without a singularity in the metric. For smooth manifolds, the manifold looks locally like the hyperspace  $\mathbb{R}^D$ in the inside and has an uniform density of nodes everywhere, therefore the argument above forbids the appearance of a condensate at any position except the boundary. A similar argument should apply to points on a smooth boundary. For this reason, we conjecture that the equation (\ref{qeqs}) for $q(x)$ always admit a solution for a smooth manifold of finite diameter with smooth boundaries and a smooth function $\pr(d)$. The argument can be easily extended to the case of a generic (non-uniform) density function $\rho(x)$, as long as $\rho(x)$ is stricty positive and smooth in all points of the manifold.  %%

Note that condensation with $\pr(d)$ an increasing (decreasing) function of the distance $d$ occurs typically in manifolds with singularities of positive (negative) curvature. An example of the latter is the above hypercone model with $\varepsilon \gg 1$ (a strongly hyperbolic space, the opposite limit with respect to a warped throat) and a decreasing function like $\pr(d)=e^{-d}$. %%

\section{Numerical results}

Since it is difficult to solve analytically the equation (\ref{qeqw}), we simulated the above model on some spaces and solved numerically the equations for $q(x)$ to compare it with the result of the simulations. Simulations were performed for two-dimensional warped manifolds with constant curvature: the flat disk (curvature $\zeta=0$), the elliptic disk (disk on the sphere, $\zeta=+1$) and the hyperbolic disk (disk in the hyperbolic plane, $\zeta=-1$). These choices of $\zeta$ are completely general since a rescaling of the curvature is equivalent to a rescaling of the radius $r_{max}$ of the disk and of the typical length $\lambda_\pr$ of the connection probability. 

We built the initial network with $n_0=2m$ nodes with $m$ links each, randomly connected. To every node, both initial and new ones, were assigned random coordinates in the manifold with a uniform distribution with respect to the volume of the manifold. The rest of the network was built as in the model presented above. 
Data reported are for simulation with total number on nodes $N=2\cdot 10^5$ and are averaged over 50 simulations with differents starting seeds.

Numerical solution of (\ref{qeqw}) is based on the numerical solution of the system of equations
\begin{align}
\frac{\partial \kb(r,t)}{\partial t}=&\frac{1}{t}(m+\kb(r,t)(\qt_1(r,t)-1)) \label{dkt} \\
\qt_1(r,t)=&\int_{r_{min}}^{r_{max}} ds\  \frac{h(r,s)m }{\int_{r_{min}}^{r_{max}} du\ h(s,u)\kb(u,t)}\label{qeqwk}
\end{align}
which is equivalent to a numerical integration of the network using macronodes. These equations are integrated up to a time $t$ when the solution satisfied the condition
\beq
\qt_1(r,t)\simeq 1-m/\kb(r,t)\equiv \qt_2(r,t)
\eeq
with a given precision. If the above condition is met, the result $\qt(r)$ is a numerical solution of (\ref{qeqw}).

\begin{figure}%%
\includegraphics[width=0.45\textwidth]{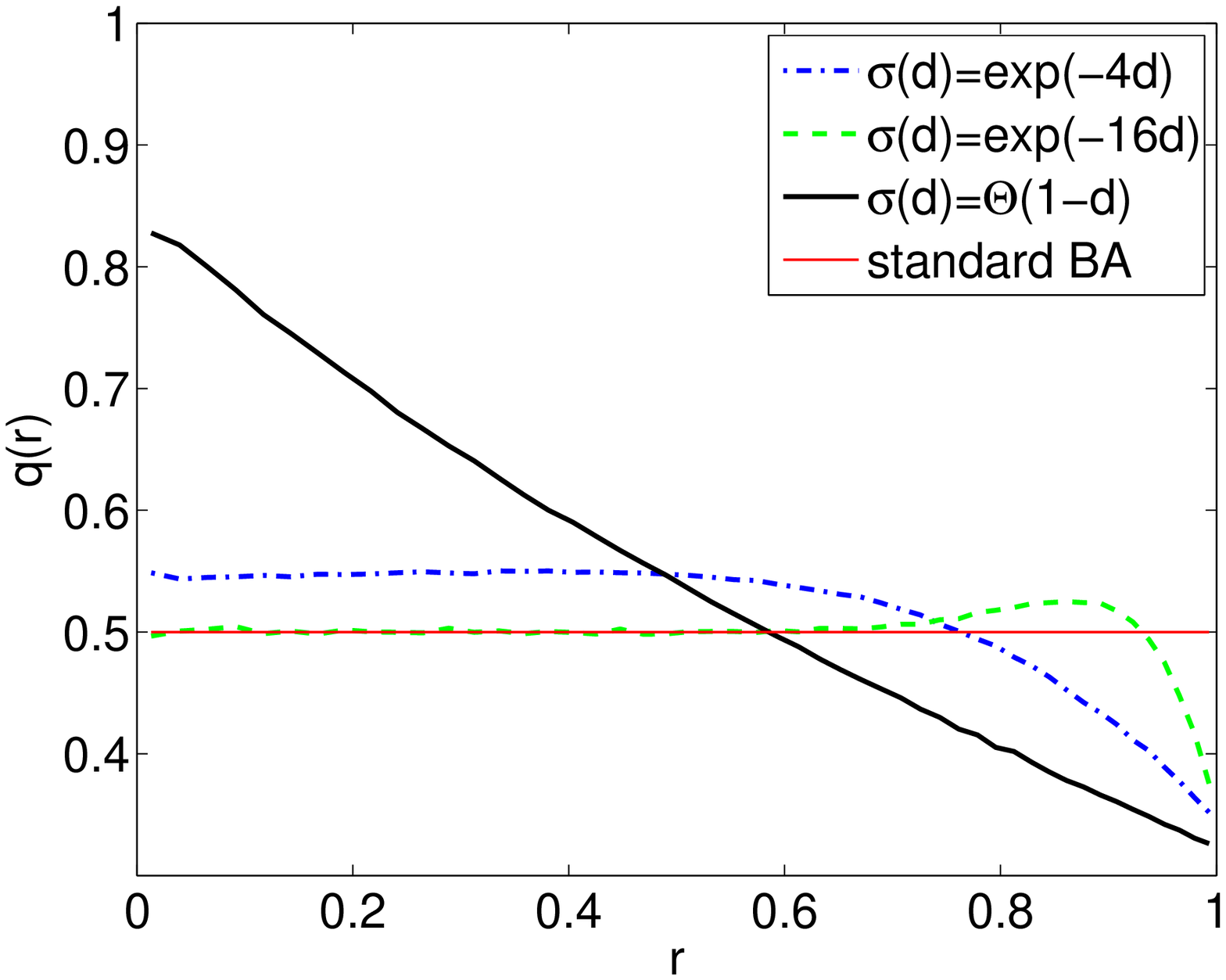}
\hspace{0.05\textwidth}
\includegraphics[width=0.45\textwidth]{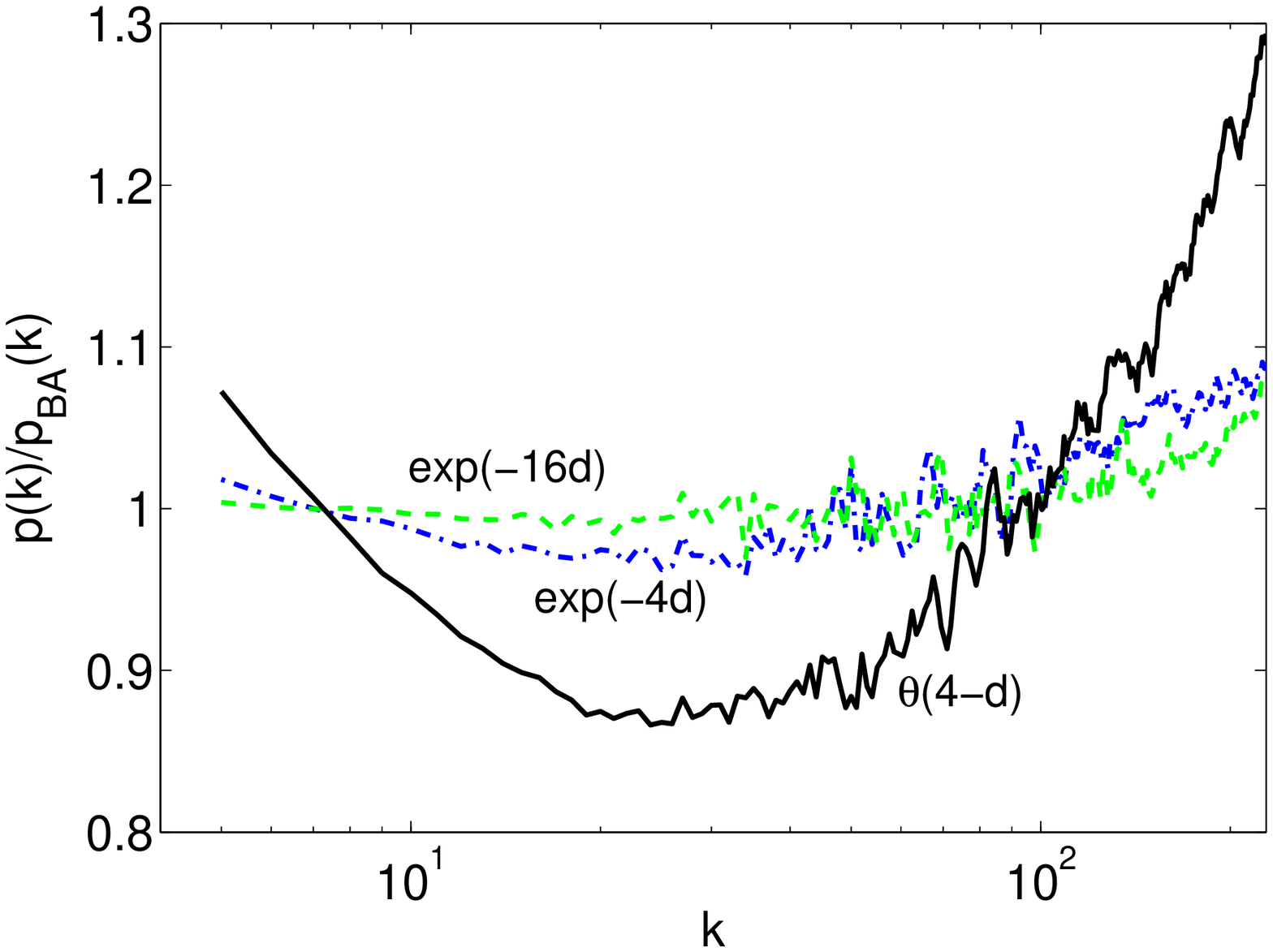}
\caption{Plots of $q(r)$ and $p(k)$ (relative to BA) for differents $\pr(d)$ for the flat disk of radius $1.5$.  }
\label{flat-integrables}	
\end{figure}

For the simulations, $q(r)$ can be estimated in two ways:
\begin{align}
\hat{q}(r)=&\frac{\log\left(\media{\frac{k_i(t_f)}{k_i(t_0)}}{r_i=r}%%
\right)}{\log(t_f)-\log(t_0)}\\
\hat{q}_2(r)=&1-\frac{m}{\media{k_i}{r_i=r}%%
}
\end{align}
where $\media{\ldots}{I}$ denotes the average over nodes in the set $I$. The difference between $\hat{q}(r)$ and $\hat{q}_2(r)$ (or $\qt_1(r,t)$ and $\qt_2(r,t)$) suggests how far the network is from equilibrium.

\begin{figure}%%
\includegraphics[width=0.45\textwidth]{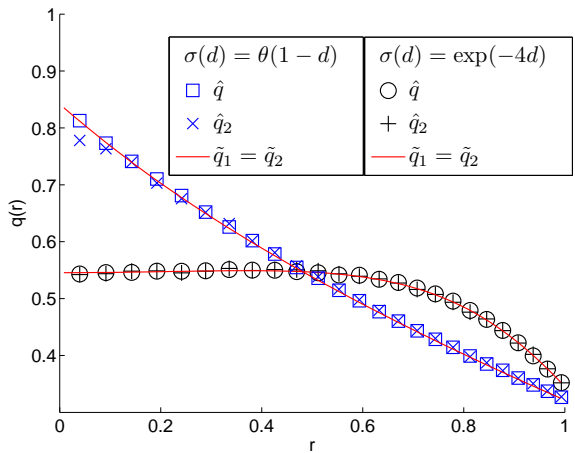}
\caption{Comparison between predicted and simulated $q(r)$ for flat disk of radius $1$ and some choices of $\pr(d)$.}
\label{simsolv}
\end{figure}

\subsection{The flat and elliptic disk}

The flat space is an homogeneous space, so the flat disk looks almost homogeneous for a short-range connection function $\pr(d)=e^{-16d/R}$ and $q(x)\simeq 0.5$ except for the region near the border, therefore the degree distribution of the network is similar to the one generated by the Barabasi-Albert (BA) model. Functions like $e^{-4d/R}$ and especially $\theta(R-d)$ have higher $\lambda_\pr$, therefore produce networks with a $q(x)$ and consequently a $p(k)$ that deviate from the BA model, as it can be seen in Figure \ref{flat-integrables}. There is a very good agreement between the numerical solution of equation (\ref{qeqw}) and $q(r)$ extracted from the simulations, %%
 as shown in Figure \ref{simsolv}. 

\begin{figure}%%
\includegraphics[width=0.45\textwidth]{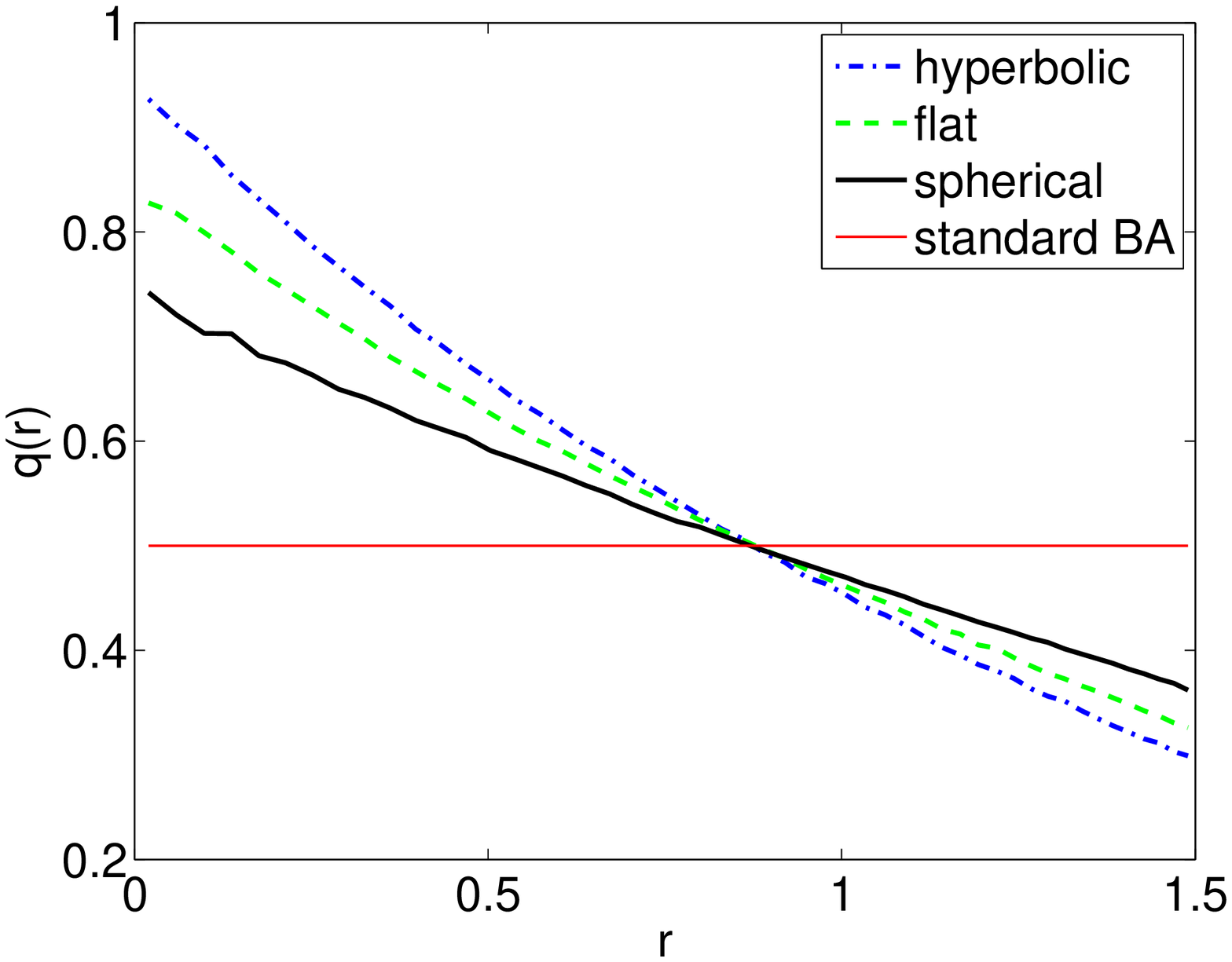}
\hspace{0.05\textwidth}
\includegraphics[width=0.45\textwidth]{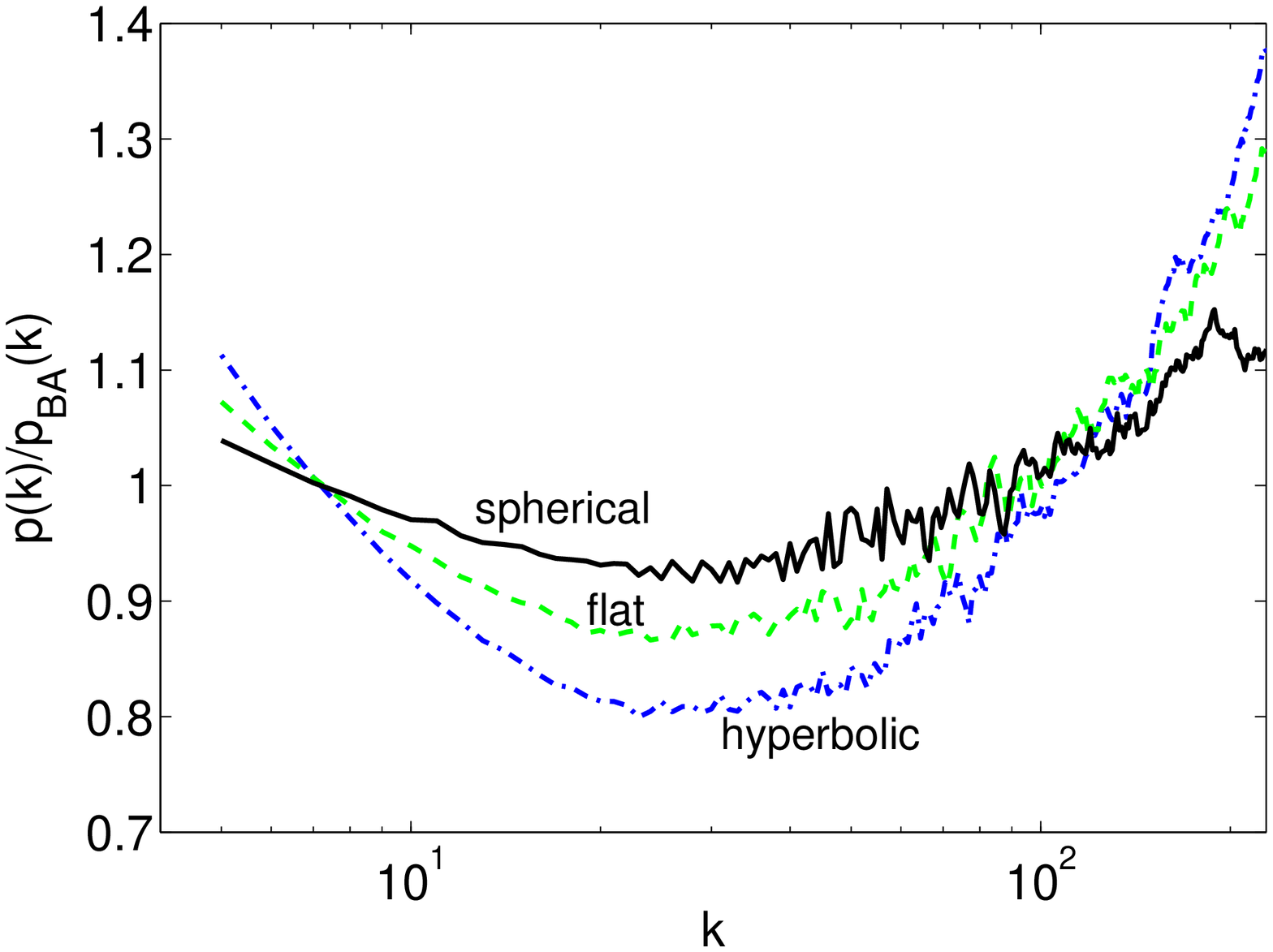}
\caption{Plots of $q(r)$ and $p(k)$ (relative to BA) for disks with $r_{max}=1.5$ and $\pr(d)=\theta(r_{max}-d)$ in two-dimensional spaces of different curvature.}
\label{figcompspaces}	
\end{figure}

Simulation results for the disk on the sphere are similar to the flat case, but the deviations from the BA case are slightly less pronounced. We compare the distributions of $q(r)$ and $p(k)$ for networks with $\sigma(d)=\theta(r_{max}-d)$ on disks in flat space, sphere and hyperbolic plane in Figure \ref{figcompspaces}. %Deviations from BA model decrease with the curvature.

\begin{figure}%%
\includegraphics[width=0.45\textwidth]{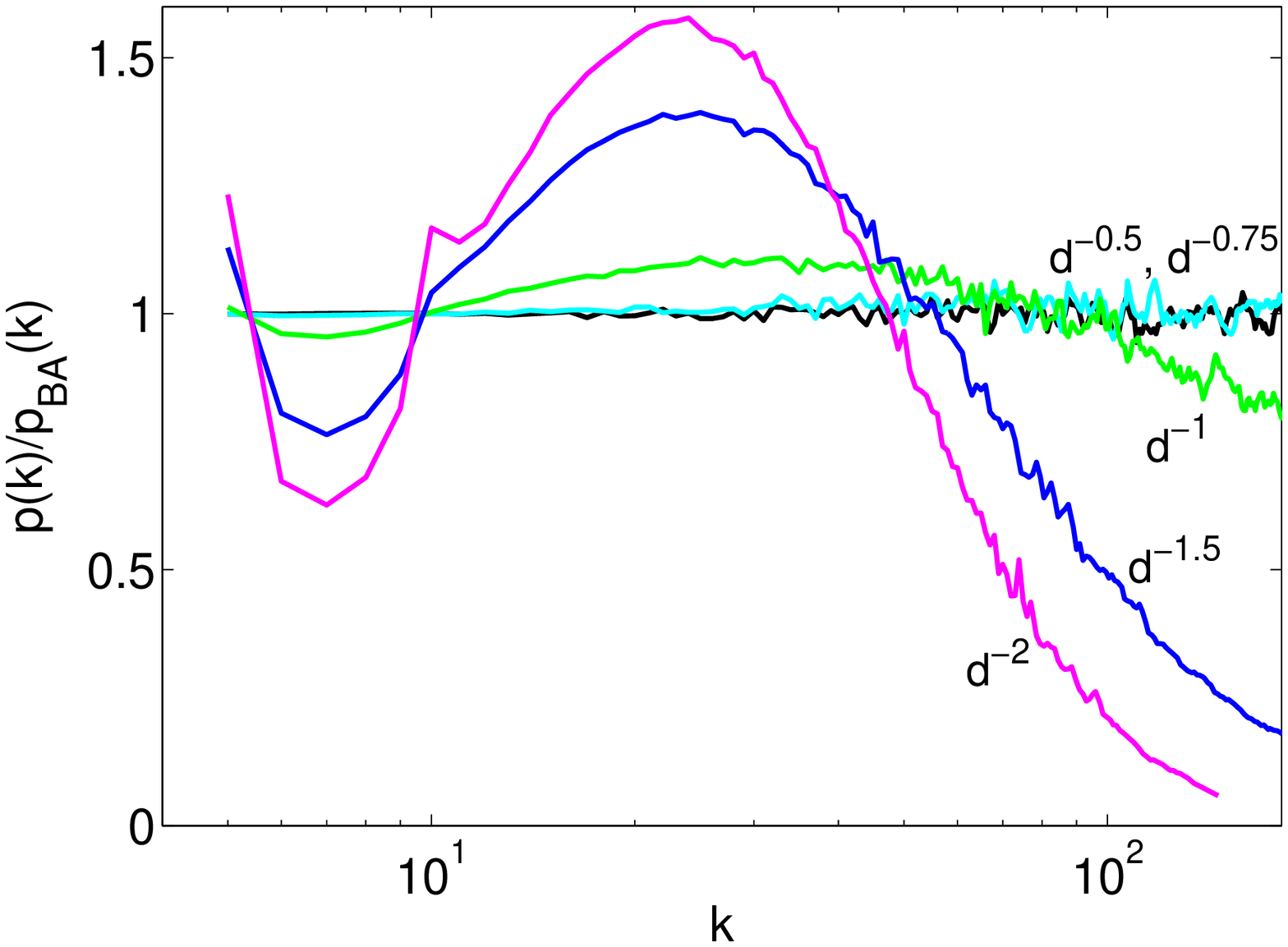}
\hspace{0.05\textwidth}
\includegraphics[width=0.45\textwidth]{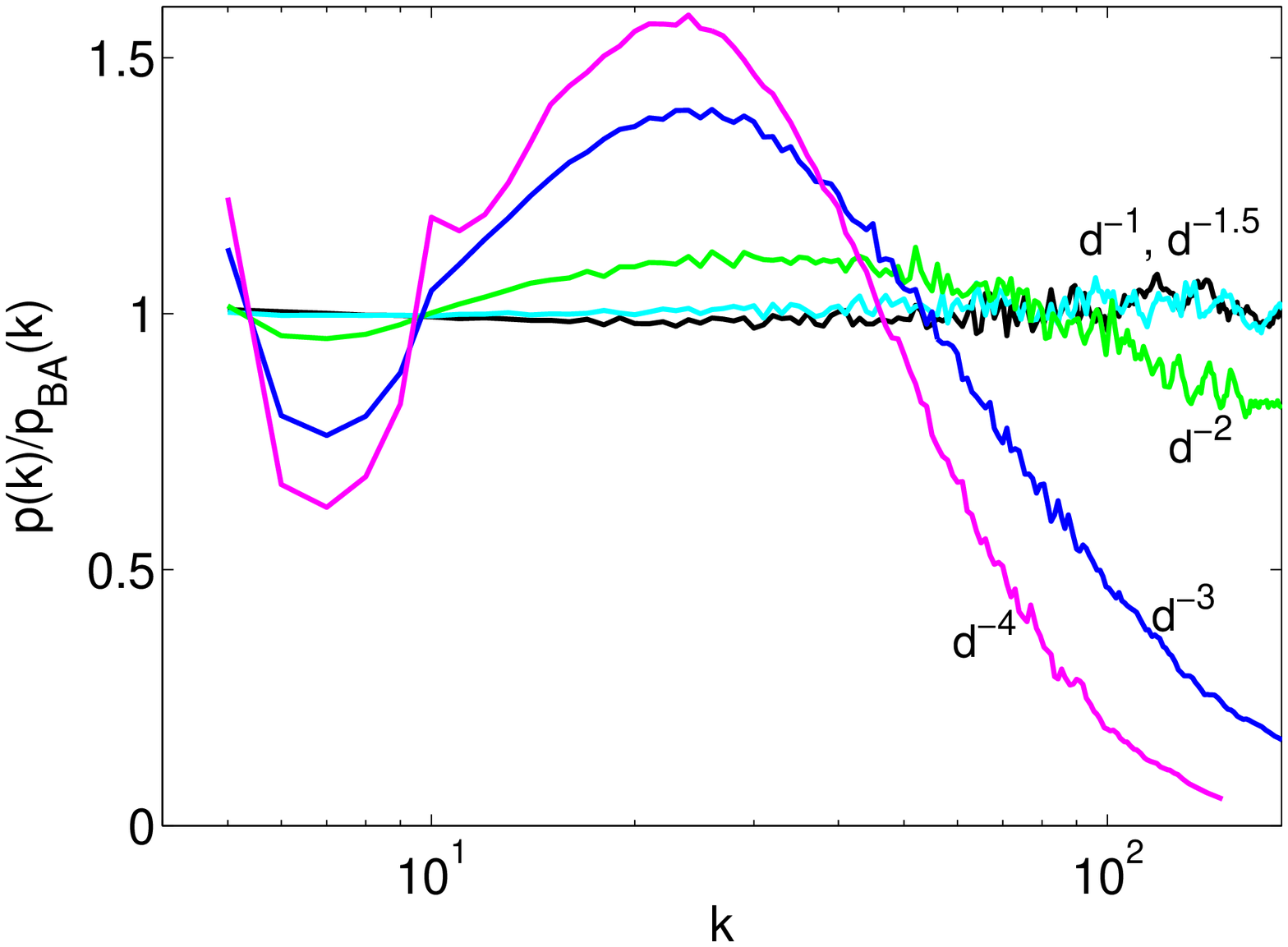}
\caption{Degree distribution (relative to BA) for the flat disk of radius 1 for different functions $\sigma(d)$ in one dimension ($\alpha_c=-1$) and two dimensions ($\alpha_c=-2$).}
\label{figdiv}	
\end{figure}

The case of networks in one- and two-dimensional flat space was actually studied in a series of works \cite{manna2002modulated,xulvi2002evolving,yook2002modeling,dellamico} with the function $\pr(d)=d^\alpha$  for negative values of $\alpha$. This case is interesting because as emphasized in \cite{xulvi2002evolving,manna2002modulated}, the degree distribution resembles a power-law for $\alpha$ above a critical value $\alpha_c$, while below this value it develops a stretched exponential tail. This transition to exponential behaviour is confirmed by our simulations for the flat disk in $D=1$ and $2$, as shown in Figure \ref{figdiv}. The critical point appears to be $\alpha_c=-1$ for $D=1$ and $\alpha_c=-2$ for $D=2$, different from the values found in \cite{manna2002modulated} for the two-dimensional torus but in agreement with later results in \cite {nandi2007transition} and the theoretical suggestions in \cite{jordan2010}.

Our approach offers a simple explanation for this behaviour. For $\alpha\leq -D$, the connection function $\pr(d)=d^\alpha$ has a divergent integral on a manifold with finite volume in dimension $D$. The rate equation approach leading to equation (\ref{qeqs}) breaks down if the integrals involving the connection function $\pr(d)$ diverge. In particular, if the divergence is localized at $d=0$, new nodes connect mostly to the nearest nodes and the distribution develops a exponential tail. The distribution resembles a stretched exponential, as shown numerically for $D=2$ in \cite{nandi2007transition} and also analytically in a similar model where each new node $i$ connects to the existing node with maximum $k_jd_{ij}^\alpha$ \cite{xie2007geographical}. Therefore the predicted critical point is $\alpha_c= -D$, in agreement with simulations in Figure \ref{figdiv}. As an interesting observation, note that the behaviour of $p(k)$ in Figure \ref{figdiv} depends apparently only on the ratio $\alpha/D$ and not on the dimension $D$ itself, at least for small $D$. 

Incidentally, the good agreement between the prediction $\alpha_c= -D$ and the simulations shown in Figure \ref{figdiv} suggests that all networks in homogeneous spaces satisfying the condition $\int_\man d^Dx\ \sigma(d(x,y))<\infty$ have the same degree distribution as the BA model and that the condition $\int_\man d^Dx\ \sigma(d(x,y))^2<\infty$ assumed in \cite{jordan2010} is not necessary.

\begin{comment}
\begin{figure}%%
\includegraphics[width=0.45\textwidth]{./confrontosphericr1q2r.eps}
\hspace{0.05\textwidth}
\includegraphics[width=0.45\textwidth]{./confrontosphericr1pk.eps}
\caption{Plots of $q(r)$ and $p(k)$ (relative to BA) for a disk with $r_{max}=1$ on the sphere for several choices of $\pr(d)$.}
\label{figsphere}	
\end{figure} 
\end{comment}

%%

%%

%%
\begin{comment}
\begin{figure}%%
\includegraphics[width=0.45\textwidth]{./confrontohypr1q2r.eps}
\hspace{0.05\textwidth}
\includegraphics[width=0.45\textwidth]{./confrontohypr1pk.eps}
\caption{Plots of $q(r)$ and $p(k)$ (relative to BA) for the hyperbolic disk with $r_{max}=1$ for several choices of $\pr(d)$.}
\label{fighypdisk}	
\end{figure} 
\end{comment}

\subsection{The hyperbolic disk}

\begin{figure}%%
\includegraphics[width=0.45\textwidth]{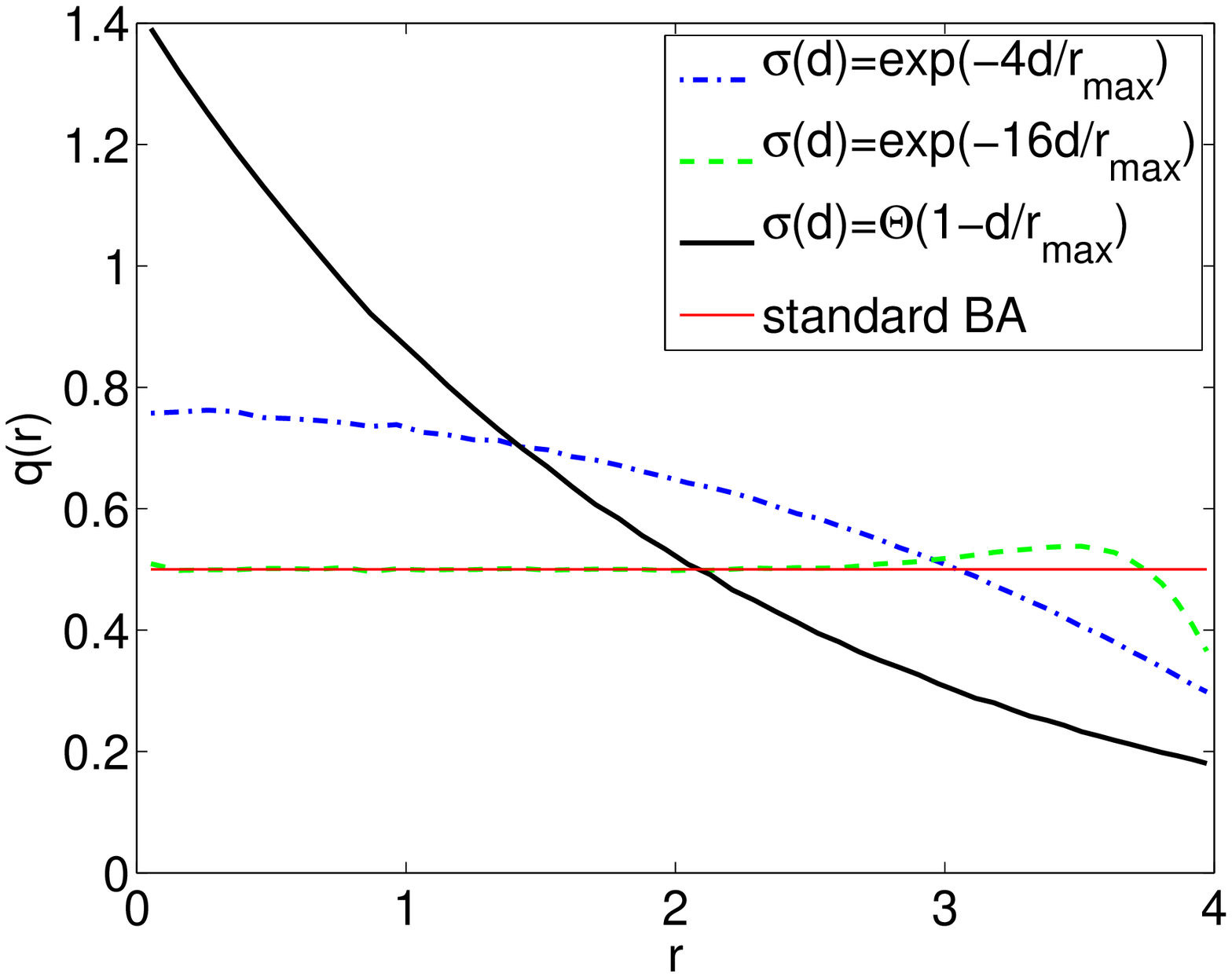}
\hspace{0.05\textwidth}
\includegraphics[width=0.45\textwidth]{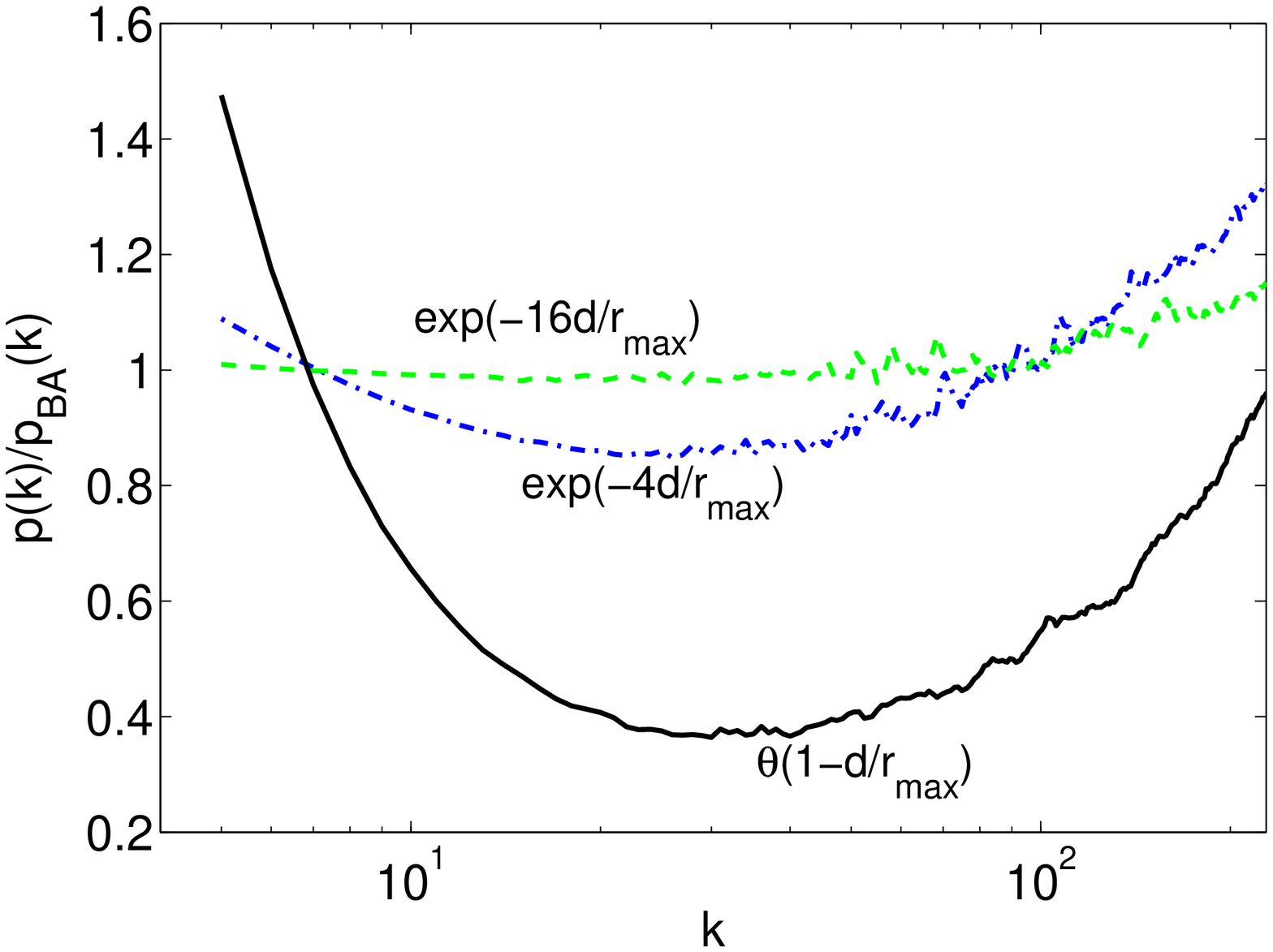}
\caption{Plots of $q(r)$ and $p(k)$ (relative to BA) for the hyperbolic disk with $r_{max}=4$ for several choices of $\pr(d)$.}
\label{fighypcond}	
\end{figure}

For disks in the hyperbolic plane of curvature -1, the results depend strongly on the radius of the disk. The deviation from the homogeneous case (BA distribution) is more pronounced as $\lambda_\pr$ and $r_{max}$ (i.e., curvature) increase. For a disk of radius $1.5$, the simulated $q(r)$ and $p(k)$ show a pattern slightly more pronounced than the flat case, while results for a disk of radius $4$ are shown in Figure \ref{fighypcond}.

The hyperbolic space is quite interesting because it resembles the scenario opposite to the warped throath discussed above, given the exponential expansion of the volumes as $r$ increase, therefore it is a good place to observe condensation in models with decreasing $\pr(d)$. Actually, the absence of singularies forbids the existence of an asymptotic condensate, but transient condensates with $q\approx 1$ could last for exponentially long times due to the exponentially small number of competing nodes near the center of the disk. This effect can be clearly seen in Figure \ref{fighypcond} with $\sigma(d)=\theta(r_{max}-d)$.

\begin{figure}%%
\includegraphics[width=0.45\textwidth]{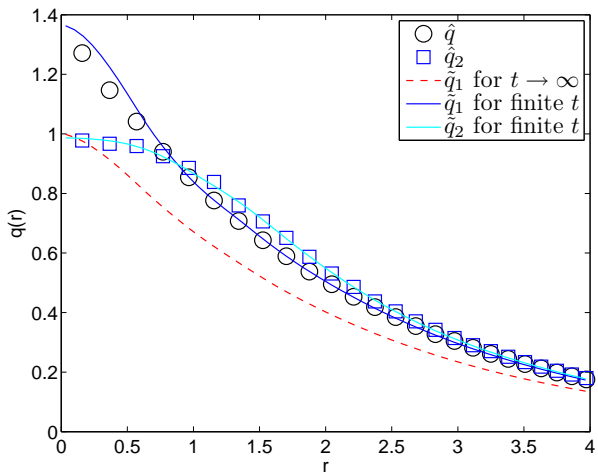}
\caption{Plots of $q(r)$ calculated numerically (both at finite and asymptotic times) and simulated for the hyperbolic disk with $r_{max}=4$.}
\label{fighypcond3}	
\end{figure}

The fact that $q(r)$ goes above 1 (that is, superlinear scaling) depends only on the long out-of-equilibrium transient regime, as it can be seen by comparing all estimators of $q(r)$ in Figure \ref{fighypcond3}. For large times, $q(0)$ is slightly below 1.
 
\section{Distribution of link lengths}
The asymptotic distribution of link lengths follows the equation 
\beq
p(l)=\int_\man d^Dx \int_\man d^Dy \frac{\pr(d(x,y))\rho(x)\rho(y)\delta(d(x,y)-l)}{\int_\man d^Dz \pr(d(z,y))\rho(z)\frac{1-q(x)}{1-q(z)}}
\eeq
which depends on the fitness $q(x)$ defined by equation (\ref{qeqs}). For (almost) homogeneous spaces, the above equation reduces to the simple distribution 
\beq
p(l)=\frac{\pr(l)\rho_d(l)}{\int_0^\infty dl' \pr(l')\rho_d(l')}\label{plength}
\eeq
where $\rho_d(l)$ is the density of nodes at distance $l$ from a random node. This result is consistent with previous studies with $\sigma(d)=d^{\alpha}$ in \cite{manna2002modulated}. 

\begin{figure}%%
\includegraphics[width=0.45\textwidth]{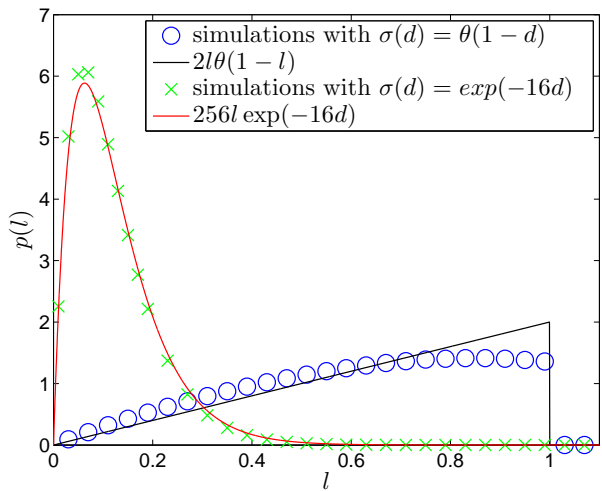}
\caption{Link length distributions $p(l)$ averaged over 50 simulations for the flat disk of radius 1 and connection probabilities $\pr(d)=\theta(1-d)$ and $e^{-16d}$. The continuous lines represent the predictions $p(l)=2l\theta(1-l)$ and $256le^{-16l}$ for homogeneous spaces. %%
}
\label{fig_dist}	
\end{figure}

Two examples of link length distributions in flat space are shown in Figure \ref{fig_dist}. The model with short-range interactions $\pr(d)=e^{-16d}$ is almost homogeneous and fits well the prediction  $p(l)=256le^{-16l}$ from equation (\ref{plength}), while the one with long-range interactions $\pr(d)=\theta(1-d)$ deviates from the distribution $p(l)=2l\theta(1-l)$ expected for an homogeneous space.  

\section{Conclusions}

The results of this paper show that the power-law behaviour of the degree distribution of growing networks with preferential attachment is quite robust with respect to the spatial structure of the network. 
The effect of the interplay of preferential attachment and the spatial nature of the network results in an heterogeneity between nodes in different positions. In spaces with high curvature or singularities, it is possible to observe Bose-Einstein condensation driven by the intrinsic heterogeneity of the space or by the finite size of the network, as it happens in hyperbolic spaces at strong curvature.

These results have been obtained through a rate equation approach, which breaks down if the integrals involved in the selfconsistency equation for $q(x)$ diverge. This is the case for connection functions $\sigma(d)$ with divergent integral. In the simple case $\sigma(d)=d^{\alpha}$, there is a critical point $\alpha_c=-D$ for the transition from the scale-free behaviour to an exponential behaviour due to distance-dominated attachment, that is, preferential connection to the nearest neighbours.

In this paper we only considered a spatial embedding of the basic Barabasi-Albert model, which gives rise to a power-law degree distribution with exponent $\gamma=3$. However, most scale-free networks found in natural and technological systems have a degree distribution with an exponent $\gamma<3$. There are several variations on the Barabasi-Albert model (see \cite{albert2002statistical} for a review) that are able to account for a different exponent and that can be included naturally in the spatial models studied in this paper without changing the main results. For example, adding $m'$ extra links per unit time would leave equation (\ref{pkspace}) invariant but change the l.h.s. of equation (\ref{qeqs}) to $q(x)/(1+2m'/m)$, which corresponds to an exponent $\gamma=2+1/(1+2m'/m)$ in homogeneous spaces, varying between $2<\gamma\leq 3$. 

Finally, spatial networks have other interesting properties, like clustering and assortativity, that have not been studied in this paper %%
and would deserve further work. In the models considered here, the volume $\mathrm{Vol}(\man)$ of the manifold is constant and the density of nodes grows linearly in time, going to infinity in the thermodynamic limit. This divergence in the density makes the clustering decrease in the thermodynamic limit. An interesting variation on these models would be a growing network with constant node density  on an expanding space. 

%\begin{comment}
\begin{acknowledgments}
We thank G. Bianconi, M. Bogu\~na, M. Mamino and S. Cremonesi for useful comments and discussions. We also thank Giuseppe Jacopo Guidi for providing hardware support. L.F. acknowledges support from CSIC (Spain) under the JAE-doc program. 
\end{acknowledgments}
%\end{comment}

\bibliographystyle{apsrev4-1}

\bibliography{networks}

\end{document}